\documentclass[conference]{IEEEtran}
\IEEEoverridecommandlockouts
\usepackage{cite}
\usepackage{amsmath,amssymb,amsfonts}
\usepackage{hyperref}
\usepackage[dvipsnames]{xcolor}

\usepackage[colorinlistoftodos,prependcaption,textsize=tiny]{todonotes}
\usepackage{microtype}
\usepackage[utf8]{inputenc}
\usepackage{graphicx}
\usepackage{textcomp}
\usepackage{xcolor}
\usepackage{subfigure}

\usepackage[noend]{algpseudocode}
\usepackage{tabularx}
\usepackage{booktabs}
\usepackage{paralist}
\usepackage{etoolbox}
\usepackage{enumitem}

\usepackage{rotating}
\usepackage{booktabs}
\usepackage{multirow}
\usepackage{verbatim }
\usepackage{booktabs,graphicx}
\usepackage[ruled,linesnumbered]{algorithm2e}

\newtheorem{prop}{Proposition}
\newtheorem{defn}{Definition}[section]

\hypersetup{
	pdftoolbar=true,        
	pdfmenubar=true,        
	pdffitwindow=false,     
	pdfstartview={FitH},    
	pdftitle={Mine Me but Don't Single Me Out: Differentially Private Event Logs for Process Mining},    
	pdfauthor={},     
	pdfsubject={},   
	pdfcreator={},   
	pdfproducer={}, 
	pdfkeywords={}, 
	pdfnewwindow=true,      
	colorlinks=true,       
	linkcolor=Brown,          
	citecolor=OliveGreen,        
	filecolor=magenta,      
	urlcolor=NavyBlue           
}

\def\BibTeX{{\rm B\kern-.05em{\sc i\kern-.025em b}\kern-.08em
    T\kern-.1667em\lower.7ex\hbox{E}\kern-.125emX}}
    
\begin{document}

\title{Mine Me but Don't Single Me Out: Differentially Private Event Logs for Process Mining}

\author{\IEEEauthorblockN{Gamal Elkoumy}
\IEEEauthorblockA{
\textit{University of Tartu}, 
Tartu, Estonia \\
gamal.elkoumy@ut.ee}
\and
\IEEEauthorblockN{Alisa Pankova}
\IEEEauthorblockA{
\textit{Cybernetica}, 
Tartu, Estonia \\
alisa.pankova@cyber.ee}
\and
\IEEEauthorblockN{Marlon Dumas}
\IEEEauthorblockA{
\textit{University of Tartu}, 
Tartu, Estonia \\
marlon.dumas@ut.ee}

}

\maketitle

\begin{abstract}
The applicability of process mining techniques hinges on the availability of event logs capturing the execution of a business process. 
In some use cases, particularly those involving customer-facing processes, these event logs may contain private information.
Data protection regulations restrict the use of such event logs for analysis purposes. 
One way of circumventing these restrictions is to anonymize the event log to the extent that no individual can be singled out using the anonymized log.
This paper addresses the problem of anonymizing an event log in order to guarantee that, upon disclosure of the anonymized log, the probability that an attacker may single out any individual represented in the original log, does not increase by more than a threshold.
The paper proposes a differentially private disclosure mechanism, which oversamples the cases in the log and adds noise to the timestamps to the extent required to achieve the above privacy guarantee.
The paper reports on an empirical evaluation of the proposed approach using 14 real-life event logs in terms of data utility loss and computational efficiency.
\end{abstract}


\vspace*{-2mm}
\section{Introduction}
\label{sec:intro}
Process Mining is a family of techniques that helps organizations 
enhance the performance, conformance, and quality of their business processes. The input of process mining techniques is an event log. An event log captures the execution of a set of instances of a process (herein called \emph{cases}). An event log consists of event records. Each record contains a reference to a case identifier, a reference to an activity, and at least one timestamp. 
Table
~\ref{tbl:state_annotated} shows an example of an event log of a healthcare process.\footnote{The columns ``Source State'' and ``Target State'' are explained later} Each case ID refers to a patient, and each activity label corresponds to a treatment or an event in the healthcare system. 

\begin{table}[hbtp]
\vspace*{-2mm}
	\centering
	\scriptsize
\caption{Event log}
\vspace*{-5mm}
 \begin{tabular}[t]{|c|c|c|c|c|}
	
\hline

Case     &     Activity     &     Timestamp      & \textit{Source State} & \textit{Target State} \\\hline

1  &  A  &  2020-08-08 10:20:0.000  & $s_0$ & $s_5$   \\
1  &  B  &  2020-08-08 10:50:0.000  & $s_5$ & $s_2$   \\
1  &  C  &  2020-08-08 16:15:0.000  & $s_2$ & $s_3$   \\
2  &  D  &  2020-08-08 12:07:0.000  & $s_0$ & $s_4$   \\
2  &  A  &  2020-08-08 13:37:0.000  & $s_4$ & $s_5$   \\
2  &  E  &  2020-08-08 14:07:0.000  & $s_5$ & $s_2$   \\
2  &  C  &  2020-08-08 19:07:0.000  & $s_2$ & $s_3$   \\
3  &  A  &  2020-08-08 13:30:0.000  & $s_0$ & $s_5$   \\
3  &  B  &  2020-08-08 13:55:0.000  & $s_5$ & $s_2$   \\
3  &  C  &  2020-08-08 20:55:0.000  & $s_2$ & $s_3$   \\
4  &  D  &  2020-08-08 15:00:0.000  & $s_0$ & $s_4$   \\
4  &  A  &  2020-08-08 17:00:0.000  & $s_4$ & $s_5$   \\
4  &  B  &  2020-08-08 17:40:0.000  & $s_5$ & $s_2$   \\
4  &  C  &  2020-08-08 23:45:0.000  & $s_2$ & $s_3$   \\
5  &  A  &  2020-08-08 16:40:0.000  & $s_0$ & $s_5$   \\
5  &  E  &  2020-08-08 17:55:0.000  & $s_5$ & $s_2$   \\
5  &  C  &  2020-08-08 23:55:0.000  & $s_2$ & $s_3$   \\ \hline
	\end{tabular}
\label{tbl:state_annotated}

\vspace*{-6mm}

\end{table}

	





Often, an event log contains private information about some individuals. 
Data regulations, such as the General Data Protection Regulation (GDPR)\footnote{\url{http://data.europa.eu/eli/reg/2016/679/oj}}, restrict the use of such event logs for analysis purposes. 
One way to overcome these restrictions is to anonymize the event log such that no individual can be singled out. 
 Singling out an individual happens when they 
 can be distinguished, within a group of people, by evaluating a predicate that discriminates him/her. The legal notion of singling out has been mathematically formalized by Cohen \& Nissim~\cite{cohen2020towards}, who define a type of attack called \emph{Predicate Singling Out (PSO) attack}.
The disclosure of the event log in Table
~\ref{tbl:state_annotated} permits singling out individuals. Specifically, the predicate ``undergoing treatment E after treatment D and A'' and the time difference between these activities can lead to a linkage attack~\cite{rafiei2020tlkc}. Given that prior knowledge, the adversary can single out the patient with case ID 2.


Privacy-Enhancing Technologies (PETs), such as differential privacy and k-anonymity~\cite{elkoumy2021privacy}, protect datasets disclosure, including event logs.
Among 
PETs, differential privacy stands out because it 
mitigates PSO attacks~\cite{cohen2020towards} and due to its composability guarantees~\cite{dwork2014algorithmic}. Differential privacy mechanisms 
inject noise into the data, quantified by a parameter called $\epsilon$. 
Lee et al.~\cite{lee2011much} show that ``the proper value of $\epsilon$ varies depending on individual values'', and that the existence of ``outliers also changes the appropriate value of $\epsilon$''. Dwork et al.~\cite{dwork2019differential} state that ``we do not know what parameter $\epsilon$ is right for any given differentially
private analysis, and we do know that the answer can vary tremendously 
based on attributes of the dataset and the policies and practices that constrain those who query it.'' Accordingly, this paper proposes a method to 
determine the needed $\epsilon$ value to disclose an event log in terms of 
a business-related metric, namely guessing advantage, which captures the 
increase in the probability that an adversary may guess information 
about an individual after the disclosure. The probability of leakage is a widely used measure of risk which can be interpreted on its own.

Usually, an adversary has prior knowledge about the individuals in the log before its release. Therefore, the adversary gains an additional advantage to guess personal information following the disclosure successfully. The goal of anonymization is to limit this risk. In this paper, we deal with the problem of limiting the additional risk by a maximum guessing advantage level $\delta$.
Specifically, we address the following problem:

\noindent\textit{ Given an event log L, and given a maximum level of acceptable guessing advantage $\delta$, generate an anonymized event log L$'$ such that the success probability of singling out an individual after publishing L$'$ does not increase by more than $\delta$.}

	

Naturally, we should ensure that the anonymized log is useful for process mining. In other words, that process mining algorithms return similar results with the anonymized log as with the original log. In this respect, a desirable property is that the anonymized log should have the same set of \emph{case variants} as the original one. A case variant is a distinct sequence of activities. For example, the case variants of the log in Table~\ref{tbl:state_annotated} are $\{ \langle A, B, C \rangle , \langle D,A, E, C \rangle , \langle D, A, B, C \rangle , \langle A, E, C \rangle \}$. This property ensures, for example, that the set of directly-follows relations between activities is not altered during anonymization. This set of relations, known as the Directly-Follows Graph, is used by automated process discovery techniques~\cite{DumasRMR18}. Moreover, the set of case variants is the main input used by conformance checking techniques. In this setting, having the same set of case variants is critical, as every case variant that is added to the log directly is a potential false positive (a deviation that does not exist in reality but it is reported as such) while every removed case variant is a potential false negative (a deviation that occurs in reality but that is not detected when using the anonymized log).
A second desirable property is that the differences between the timestamps of consecutive events in the anonymized log are as close as possible to those in the original log, as these time differences are used by performance mining techniques~\cite{DumasRMR18}.
Accordingly, we tackle the above problem subject to the following requirements:
\begin{enumerate}[label=\textbf{R\arabic*}]

	\item\label{int:req:trace}
	The anonymized event log must have the same set of case variants as the original log. 
	
	
	 
	\item\label{int:req:time}
	The difference between the real and the anonymized time values is minimal given the risk metric $\delta$. 

\end{enumerate}
The first requirement should be tackled in a way that keeps the variant frequencies and the log size close to the original log.
The second requirement can be tackled w.r.t. different attack models. In this paper, we consider an attack model wherein the attacker seeks to learn a prefix or a suffix of an individual's trace or the timing of a given activity for a given individual.

In this paper, we assume that the activity labels are public.
Under this starting point, we tackle the above problem by defining a notion of the differentially private event log. Given a maximum allowed guessing advantage, $\delta$, a differentially private event log is obtained by oversampling the traces in the log and injecting noise to the event timestamps. This ensures that the probability that an attacker may single out any individual, based on the prefixes/suffixes of the individual's trace or based on the event timestamps, is not more than $\delta$.


The proposal relies on a data structure that compactly captures all prefixes and suffixes of a set of traces, namely a Deterministic Acyclic Finite State Automata (DAFSA)~\cite{daciuk2000incremental}. A DAFSA is a lossless representation of an event log, wherein every prefix or suffix shared by multiple traces is represented once. 
By analyzing the frequency and time differences of each DAFSA transition, we determine the amount of oversampling and timestamp noise. The estimated noise limits the guessing advantage 
of an attacker by inspecting the traces 
traverse this transition in the anonymized log. 



The paper is structured as follows. Sect.~\ref{sec:background} introduces background notions and related work. Sect~\ref{sec:preliminaries} presents the notion of the differentially private event log, the attack model, and the risk quantification approach. Sect.~\ref{sec:approach} translates the risk quantification into an algorithm that anonymizes a log. Finally, Sect.~\ref{sec:eval} presents an empirical evaluation while Sect.~\ref{sec:conclusion} draws conclusions and discusses future work.

\section{Background and Related Work}
\label{sec:background}

\subsection{Differential Privacy}
\label{sec:diffPriv}



A database $D$ is a set of attributes with values drawn from a universe $U$. A tuple is an instance in the database, which is a collection of values of a set of attributes $A= A_1, A_2, ..., A_m$, where $m$ is the number of attributes in  $D$. A tuple corresponds to an individual who requires their privacy to be maintained. An individual may contribute to more than one tuple.

A mechanism $M: D \rightarrow Range (M)$ maps a database $D$ to a particular distribution of values $Range(M)$ (e.g., to a vector of real numbers). A privacy mechanism $M$ can be either unbounded or bounded $\epsilon$-differentially private ($\epsilon$-DP). An unbounded $\epsilon$-DP mechanism makes it hard to distinguish two databases that differ in the \emph{presence} of one tuple~\cite{dwork2014algorithmic}. On the other hand, a bounded $\epsilon$-DP mechanism makes it hard to distinguish two databases that differ in the \emph{value} of one tuple.

\begin{defn} [Unbounded $\epsilon$-differentially private mechanism~\cite{dwork2014algorithmic}]\label{def:udp}
A mechanism $M$ is said to be $\epsilon$-differentially private if, for all the data sets $D_1$ and $D_2$ differing \underline{at most on one item}, and all $S \subseteq Range (M)$, we have $Pr[M(D_1) \in S] \leq exp(\epsilon) \times Pr[M(D_2) \in S]$.

\end{defn}

\begin{defn} [Bounded $\epsilon$-differentially private mechanism~\cite{dwork2014algorithmic}]\label{def:bdp}
A mechanism $M$ is said to be $\epsilon$-differentially private if, for all the data sets $D_1$ and $D_2$ differing \underline{at most on the value of one item}, and all $S \subseteq Range (M)$, we have $Pr[M(D_1) \in S] \leq exp(\epsilon) \times Pr[M(D_2) \in S]$.

\end{defn}

In some cases, it is desired to apply differential privacy to only values of a particular attribute $A$, e.g., the attribute timestamp in Table
~\ref{tbl:state_annotated}. 
We apply differential privacy to $D_1$ and $D_2$ w.r.t the attribute $A$, i.e., $D_1$ and $D_2$ differ only on $A$'s value in a single tuple. Moreover, we want to take into account the particular \emph{amount of change} in attribute $A$. 

\begin{defn} [Bounded $\epsilon$-differentially private mechanism w.r.t. attribute]\label{def:bdp_attribute}
A mechanism $M$ satisfies $\epsilon$-differential privacy if all the data sets $D_1$ and $D_2$ differing at most on the value of one attribute $A$ for one tuple in $D_1$ and $D_2$, and all $S \subseteq Range (M)$, we have $Pr[M(D_1) \in S] \leq exp(\epsilon \cdot |D_1.A - D_2.A|) \times Pr[M(D_2) \in S]$.

\end{defn}

The $\epsilon$-differential privacy restricts the ability to single out an individual (Def.~\ref{def:udp} and \ref{def:bdp}) or disclose an individual's private attribute (Def.~\ref{def:bdp_attribute}).
We consider an interactive mechanism~\cite{dwork2014algorithmic}, where a user submits a query function $f$ to a database and receives a noisified result. Formally, there is a mechanism  $M_f$ that computes $f$ and injects noise into the result. The amount of 
noise depends on the \textit{sensitivity} of $f$, which quantifies how much change in the input of $f$ affects change in its output.

	



\begin{defn} [Global Sensitivity] \label{def:gs2}
Let $f : D \rightarrow \mathbb{R}^d$.
\begin{itemize}
\vspace*{-1.5mm}
\item Global sensitivity w.r.t. presence of a tuple is $\Delta f= \max\limits_{D_1,D_2} |f(D_1) - f(D_2)|$;
\item Global sensitivity w.r.t. attribute $A$ is $\Delta^A f= \max\limits_{D_1,D_2} \frac{|f(D_1) - f(D_2)|}{|D_1.A - D_2.A|}$;
\end{itemize}
\vspace*{-1mm}
where $\max$ is computed over all databases $D_1,D_2 $ differing in one item at most.
\end{defn}

Given the database $D$ and the query function $f$, a randomized mechanism $M_f$ returns a noisified output $f(D)+Y$, where $Y$ is a value drawn randomly from a particular distribution. E.g., we can draw values from a Laplace distribution $Lap(\lambda, \mu)$, which has a probability density function $\frac{1}{2\lambda} exp(-\frac{|x-\mu|}{\lambda})$, where $\lambda$ is a scale factor, and $\mu$ is the mean. It is known~\cite{dwork2014algorithmic} that, for real-valued $f$, if we set $\mu=0$, for $\lambda = \frac{\Delta f}{\epsilon}$ we obtain an $\epsilon$-DP mechanism w.r.t. row presence (Def.~\ref{def:udp}), and for $\lambda = \frac{\Delta^A f}{\epsilon}$, we obtain an $\epsilon$-DP mechanism w.r.t. attribute $A$ (Def.~\ref{def:bdp_attribute}).






 The privacy parameter 
 $\epsilon$ ranges from $0$ to $\infty$, and the desired level of privacy 
 depends on the data distribution. 
 This brings us to the question of how much $\epsilon$ is enough~\cite{lee2011much}? Laud et al.~\cite{laud2020framework} propose a framework to quantify $\epsilon$ from a probability value called the \textit{guessing advantage}, which is the increase in the probability that an adversary reveals the value after the data disclosure. In this paper, we adopt the work by Laud et al.~\cite{laud2020framework} to provide a 
 quantification of $\epsilon$ based on the guessing advantage threshold, with the assumption that an adversary has background knowledge about all the other instances.


\vspace*{-1mm}
\subsection{Privacy-Preserving Process Mining}
\label{sec:pppm}
\vspace*{-1mm}
The use of PETs for privacy-preserving process mining (PPPM) has been considered in previous studies~\cite{elkoumy2021privacy}. 
PRETSA~\cite{fahrenkrog2019pretsa} and TLKC~\cite{rafiei2020tlkc} propose k-anonymity mechanisms to anonymize event logs. 
These studies do not fulfill~\ref{int:req:trace} because the case variants in the anonymized log are different from the original log due to the suppression of cases or events within cases. In one example in~\cite{rafiei2020tlkc}, TLKC suppressed 87\% of the activities in the output.
Furthermore, k-anonymity does not fully mitigate PSO attacks~\cite{cohen2020towards}.
Mannhardt et al.~\cite{mannhardt2019privacy}  define a differential privacy mechanism to anonymize two types of queries: the query ``frequencies of directly-follows relations'' and ``frequencies of trace variants''. 
PRIPEL~\cite{fahrenkrog2020pripel} propose timestamp shifts to anonymize the timestamp attribute of the event log. 
These two studies do not address the problem stated in Sect.~\ref{sec:intro} because they do not limit the guessing advantage to a certain threshold. Instead, the user has to provide an $\epsilon$ value as an input. Furthermore, these approaches do not fulfill~\ref{int:req:trace}, as the set of case variants in the anonymized log is different from the original.




Other related work of 
PPPM is orthogonal to our research.
For example, privacy quantification methods have been proposed in~\cite{von2020quantifying} and~\cite{rafiei2020towards}.
Batista \& Solanas~\cite{batista2021uniformization} present an approach that anonymizes event logs w.r.t. resources.
Furthermore, other studies addressed the 
secure processing of distributed event logs~\cite{elkoumy2020secure,elkoumy2020shareprom}.





\section{Differentially Private Event Log}
\label{sec:preliminaries}
In this section, we introduce the concepts of event log and DAFSA. We then define a differentially private event log notion, and we present our attack model and risk quantification.
\subsection{Event Log Representation}
\label{sec:event_log_rep}
An event log is a set of event records (events for short) capturing the execution of activities of a process. Each event contains a unique identifier of the process instance in which it occurs (case ID), an activity label, and a timestamp. In addition, an event may contain other attributes, e.g., resources. 
This paper focuses on anonymizing three attributes: case ID, activity label, and timestamp.
If we group the events in an event log by case ID, and we sort the events in each group chronologically, then every resulting group is called a trace. A trace captures the sequence of events that occurred in a case.

\begin{defn} [Event Log, Event, Trace]\label{def:event_log}
	An event log $L= \{e_1, e_2, ..., e_n\}$ of a process is a set of events $e=(i,a,ts)$, each capturing an execution of an activity $a$, with a timestamp $ts$, as part of a case $i$ of the process. The trace $t=\langle e_1, e_2, ..., e_m\rangle$ of a case $i$ is the sequence of all events in $L$ with identifier $i$, ordered by timestamp.
\end{defn}

An event log $L$ may be represented as a set of traces $\{ t_1, t_2, ..., t_k \}$. If we set aside the case ID and the timestamp of a trace and focus on the activity labels, we can represent a trace as a word over the alphabet of activity labels. Each particular word extracted from a log in this way is called a \emph{case variant} of the log.

\begin{defn}[Case Variant] 
Given an event log $L$, a case variant of $L$ is a sequence of activity labels $\langle\, a_1, a_2, ..., a_k \,\rangle$ such that there is a trace $\langle\, e_1, e_2, ..., e_k \,\rangle$ of $L$ such that $\forall j \in [1..k] \; id(e_j) = v_j$, where $id(e)$ is the case ID of the event $e$.
\end{defn}



We assume that each trace in a log corresponds to the execution of the process pertaining to an individual whose privacy we wish to safeguard. Specifically, our goal is to mitigate singling out of an individual based on any prefix or suffix of their trace. To achieve this, we group the prefixes and suffixes in the log, and we inject independent differentially private noise to each such group. For this, we need a representation of the log that partitions the prefixes and suffixes of the log traces into groups. In other words, this representation should assign each prefix (suffix) in the log to one group such that the union of the groups is equal to the entire set of prefixes (suffixes). In addition, we also require that this representation preserves the set of case variants of the log (cf.~\ref{int:req:trace}).

The DAFSA provides such a partitioning. Given a set of words, each state in the DAFSA represents a group of prefixes that share the same set of suffixes, and suffixes that share the same set of prefixes~\cite{daciuk2000incremental}. 
An advantage of the DAFSA over similar representations, such as 
tries, is that a DAFSA contains a minimal number of groups (states)~\cite{daciuk2000incremental}. By minimizing the number of groups, we obtain larger groups.
The larger the group, the smaller the noise generated by differential privacy.

\begin{figure}[htbp]
  \vspace*{-4mm}
  \centering
\includegraphics[width=.70\columnwidth]{{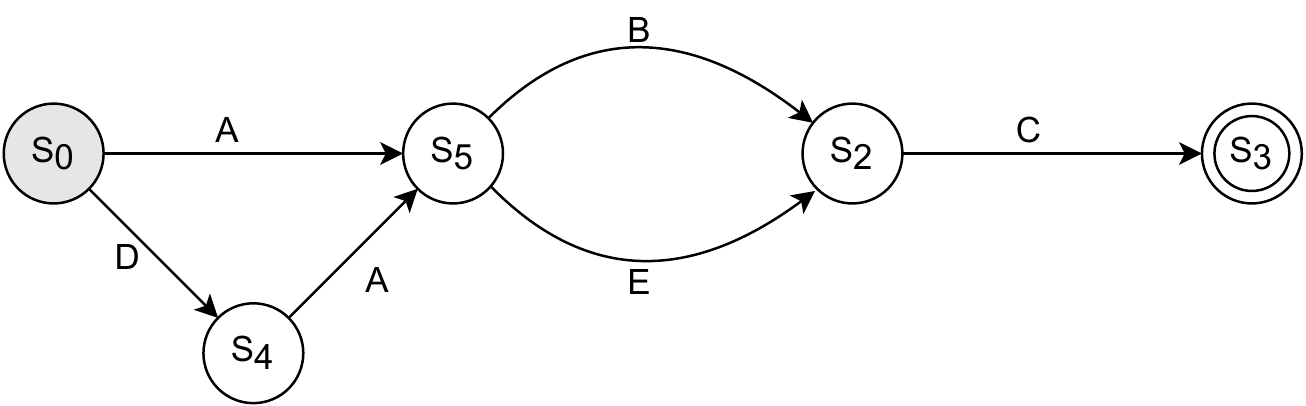}}
\vspace*{-4mm}
	\caption{ DAFSA of the event log in Table~\ref{tbl:state_annotated}}
	\label{fig:dafsa}
  \vspace*{-3mm}
\end{figure}

Given a DAFSA constructed from a set of words, every word is a path from an initial state to a final state. Conversely, every path from an initial state to a final state corresponds to a word in the given set of words~\cite{daciuk2000incremental}.
Reissner et al.~\cite{reissner2017scalable} reuse the algorithm in~\cite{daciuk2000incremental} to represent a log as a DAFSA. Every trace is seen as a word (its corresponding case variant).
For example, the DAFSA of the log in Table~\ref{tbl:state_annotated} is shown in Fig.~\ref{fig:dafsa}. 

\begin{defn}[DAFSA~\cite{daciuk2000incremental}]\label{def:dafsa}
Let $V$ be a finite non-empty set of (activity) labels. A DAFSA is an acyclic and deterministic directed graph $D = ( S, s_0, A, S_f)$, where $S$ is a finite set of states, $s_0 \in S$ is the initial state, with a set of arcs $A \subset S \times V \times S$, and a set of final states $S_f$.
\end{defn}




The common prefixes of the DAFSA in Fig.~\ref{fig:dafsa} are $\{\langle A,B\rangle\, \langle D,A\rangle\, \langle A\rangle \}$, and the common suffixes are $\{\langle B,C\rangle\, \langle E,C\rangle \}$.
Cases corresponding to case variants that traverse a given DAFSA state $s$ share the same 
set of prefixes and suffixes. In this paper, we employ DAFSA states and transitions 
to group common 
prefixes and suffixes within cases of the log. We annotate the log with the DAFSA transitions and states to relate such grouping to the event log cases.

\begin{defn} [State Annotated Event Log]
\label{def:state_el}
A state annotated event log $L_s =\{r_1, r_2, ..., r_n\}$ is a set of entries $r=(i,a,ts, s_i, s_e)$, each links an event $e=(i,a,ts) \in L$, where $L$ is the event log, to the DAFSA transition $t=(s_i, a, s_e)$ that represents the occurrence of that event, where $t$ starts from $s_i$, ends at $s_e$, and labeled with the same activity $a$.

\end{defn}

The state-annotated event log links every event in the event log, based on its prefix, suffix, and activity label, to the DAFSA transition. 
Every event is labeled by the source state and target state of the DAFSA transition. 
Table~\ref{tbl:state_annotated} (columns source state and target state) shows the state annotated event log.

	




\subsection{Privacy Mechanism}
\label{sec:diffprivEL}

Given 
Definitions~\ref{def:dafsa}, and ~\ref{def:state_el}, all cases that share a common prefix will traverse a given state $s$ in the DAFSA corresponding to this prefix. The same holds for cases that share a common suffix. 
We quantify the privacy parameter $\epsilon$ using the state annotated event log to mitigate singling out an individual by their prefixes (or suffixes). To study the histogram (count) distribution of the common prefixes and suffixes between traces of an event log, we construct the \textit{DAFSA transition contingency table}. A contingency table is a histogram of group occurrences. 

\begin{defn} [DAFSA Transitions Contingency Table]
The DAFSA transition contingency table $C$ is the histogram of counts for each transition $t=(s_i, a, s_e)$ of the DAFSA $D$, where $s_i$ is the source state, $a$ is the activity label, and $s_e$ is the target state of $t$.
\end{defn}
\begin{table}[hbtp]
\vspace*{-5mm}

\centering
\scriptsize
\caption{DAFSA Transitions Contingency Table}
\vspace*{-5.5mm}
    \begin{tabular}[t]{|c|c|c|c|}
        
        \hline
        
        \shortstack{Source\\State}     &     Activity     &     \shortstack{Target\\State}      & Count \\\hline
        
        $s_0$  &  A  &  $s_5$  & 3    \\\hline
        $s_5$  &  B  &  $s_2$  & 3    \\\hline
        $s_2$  &  C  &  $s_3$  & 5    \\\hline
        $s_0$  &  D  &  $s_4$  & 2    \\\hline
        $s_4$  &  A  &  $s_5$  & 2    \\\hline
        $s_5$  &  E  &  $s_2$  & 2    \\
         \hline
    \end{tabular}

\label{tbl:dafsa_contingency}
\vspace*{-5.5mm}
\end{table}
Table~\ref{tbl:dafsa_contingency} shows the DAFSA transitions contingency table. These counts are usually called marginals. The marginals contain the correlations counts of the common sets of prefixes and suffixes of the DAFSA. 
We anonymize these marginals to prevent singling out that an individual has been through an activity, using the prefix and the suffix set of activities. 
The anonymization of the marginals can be directly mapped to the privacy-preserving Online Analytical Processing~\cite{barak2007privacy}.

\begin{defn} [differentially private DAFSA Transitions Contingency Table]
\label{def:dp_dafsa}
Let $f$ be a query function that computes a DAFSA transitions contingency table that has a set of transitions $t=(s_i, a, s_e)$ and a count cell $c_i$ for each transition. Let $M_f$ be an unbounded $\epsilon$-differentially private mechanism (by Def.~\ref{def:udp}) that injects noise into the result of $f$. The differentially private DAFSA transitions contingency table is: 
$M(C):=\{(t_1,M_f(c_1)),(t_2,M_f(c_2)) ...,(t_n,M_f(c_n))\}$, where $n$ is the number of DAFSA transitions.
\end{defn}


Our 
goal is not to anonymize a DAFSA but an \emph{event log}. 
We assume that case IDs have been pseudonymized, the activity labels are public, and the individual cases are independent. We propose a mechanism $M$ that anonymizes two properties of an event log: the activity timestamp and the set of prefixes and suffixes of activity. 
We can release the time for the first property by applying a bounded $\epsilon$-DP mechanism w.r.t. timestamp attribute (Def.~\ref{def:bdp_attribute}). We need a mechanism that makes the contingency table $\epsilon$-DP (Def.~\ref{def:dp_dafsa}) for the second property.


\begin{defn}[differentially private Event Log]
\label{def:diff_priv_EL}
Let $L$ be an event log as defined in Def.~\ref{def:event_log}. We say that a log $M(L)$ is $\epsilon$-differentially private if: (1) it is $\epsilon$-differentially private w.r.t. timestamp (Def.~\ref{def:bdp_attribute}); and (2) its lossless representation is $\epsilon$-differentially private (Def.~\ref{def:dp_dafsa}).

\end{defn}



	




The mechanism $M_f$ of Def.~\ref{def:dp_dafsa} operates on counts and cannot be applied to a log directly. We need
to translate the noise injection of the contingency table to the log.
Kifer et al.~\cite{kifer2011no} present the notion of a \textit{move} as a way of anonymizing the marginals of contingency tables. A move is a process that adds or deletes a tuple from the contingency table. 
In order to fulfill the requirement~\ref{int:req:trace}, we define \textit{oversampling} as increasing a count (positive move) in the contingency table by replicating a random tuple in the log. 

\begin{defn} [DAFSA Transition Oversample]
\label{def:oversampling}
Given a DAFSA transition contingency table $C_i$, an oversample $O$ is a transformation that adds a  DAFSA transition instance to $C_i$, producing a contingency table $C_j=O(C_i)$, with an increase of only one count cell by 1. 

\end{defn}

We define a mechanism $M$ that oversamples tuples of a log $L$ so that, if we computed the contingency table of $M(L)$, it would be $\epsilon$-DP. We must be careful that $M(L)$ does not leak anything that the contingency table would not leak. For example, if we keep timestamps unchanged, it suffices to remove the tuples with repeating timestamps to eliminate duplicates, so we need to make the times $\epsilon$-DP as well. Since duplicated timestamps are correlated, we have to divide the $\epsilon$ values of the time queries of the replicated cases by their oversampling ratio. 
To keep the counts in the DAFSA transitions consistent, we oversample the prefix and the suffix of the oversampled transition, i.e.\ we always oversample entire cases.

\begin{defn} [Case Oversample]
\label{def:cases_oversampling}
Given an event log $L$, a case oversample $O_c$ is a transformation that duplicates a case $c_i$ of log $L$, in such a way that $c_j$, the duplicated case in log $O_c(L)$, and $c_i$ have the same sequence of activities. 

\end{defn}



\subsection{Risk Quantification and Attack Model}
\label{sec:attack_model}
Given Def.~\ref{def:diff_priv_EL}, we seek to quantify $\epsilon$'s value to publish a differentially private log. We consider a scenario where an organization shares its event log with an analyst, who should not be able to infer an individual's information. The analyst has background knowledge about all the individuals in the 
log except for the individual of their interest.
Specifically, we consider the case where an attacker (the analyst) uses the event log to single out an individual. The attacker has a goal $h(L)$ that captures 
their interest in an event log $L$. We specifically consider the following attacker's goals:
\begin{itemize}
    \item $h_1$: Has the individual been through a specific subtrace (prefix or suffix)? The output is a bit with a value $\in \{0,1\}$ that represents yes or no.
    \item $h_2$: What is the execution time of a particular activity that has been executed for the individual? The output is a real value to be guessed with precision.
\end{itemize}



We use an $\epsilon$-differential privacy mechanism to mitigate the above attacker's goals. To determine the privacy parameter $\epsilon$, we adopt the guessing advantage framework~\cite{laud2020framework}, which quantifies the risk of publishing a dataset as the difference between two probabilities: the prior and the posterior guessing probability.
Even without publishing the event log, attackers can use their knowledge to guess information about a specific individual. 
The guess is considered successful if it falls within a range of values $H_p$, which is the actual value $\pm$ a precision.

\begin{defn}[Prior Guessing Probability]\label{def:prior}
An attacker's prior guessing probability is defined as $P := Pr[h(L) \in H_p]$


\end{defn}




A guessing precision $p$ is a percentage value representing the range of a successful guess $H_p$. For example, if the true value is 0.5 and $p= 0.2$, the guessed value is 
successful if it falls in range $H_p= [0.3..0.7]$. Our method pre-processes the log to normalize the 
values (relative timestamps) to be between 0 and 1. The precision is interpreted within this [0,1] range.

The guessing advantage is 
defined as the difference between the posterior probability (after publishing $M(L)$) and the prior probability (before publishing $M(L)$) of an attacker making a successful guess in $H_p$. Let $\delta$ be the maximum allowed guessing advantage, stated by the event log publisher.




\begin{defn}[Guessing Advantage]\label{def:ga}
Attacker's advantage in achieving the goal $h$ with precision $p$ is at most $\delta$ if, for any published event log $L'$, 
\vspace*{-3mm}
\[Pr[h(L') \in H_p\ |\ M(L)=L'] - Pr[h(L) \in H_p] \leq \delta.\]
\vspace*{-7mm}
\end{defn}

Laud et al.~\cite{laud2020framework} quantify the maximum value of $\epsilon$, that achieves the upper bound $\delta$.

\begin{prop}\label{prop:min_eps}
The maximum possible $\epsilon$ (i.e., the minimum noise) that achieves the upper bound $\delta$, w.r.t. the above attack model, and fulfills the requirement~\ref{int:req:time} is
\vspace*{-2mm}
\begin{equation}\label{eqn:epsilon_time}
\epsilon_{k}= - \ln\left( \frac{P_{k}}{1-P_{k}}(\frac{1}{\delta+ P_{k}} -1)\right) \cdot \frac{1}{r}\enspace,
\end{equation}
\vspace*{-1mm}
where $r$ is the maximum value in the range of values ($r=1$ after the normalization of values), and $P_k$ is the prior guessing probability for an instance $k$.
\end{prop}
The proof of Prop.~\ref{prop:min_eps} is available in the supplementary material~\cite{supplementary}, where it appears with heading ``Proposition 2''.


Given Prop.~\ref{prop:min_eps}, if there is enough data, we can estimate $P_k$ based on the data distribution as:
\vspace*{-1mm}
\begin{equation}\label{eqn:p_k}
P_{k} = CDF(t_{k} + p \cdot r) - CDF(t_{k} - p \cdot r)\enspace,
\end{equation}
\vspace*{-5mm}

where $t_k$ is the value of 
an instance, and $p$ is the precision.
Suppose we cannot estimate the probability distribution of input values (e.g., the likelihood of participating in a subtrace). In that case, we can compute the worst-case scenario $P_k$ as~\cite{laud2020framework}:
\vspace*{-2mm}
\begin{equation}\label{eqn:p_freq}
P_k=(1-\delta) / 2\text{ for all $k$}\enspace.
\vspace*{-1mm}
\end{equation}






\section{Computing Differentially Private Event Logs}
\label{sec:approach}

\begin{figure}[htbp]
  \vspace*{-4mm}
\centerline{\includegraphics[width=.98\columnwidth]{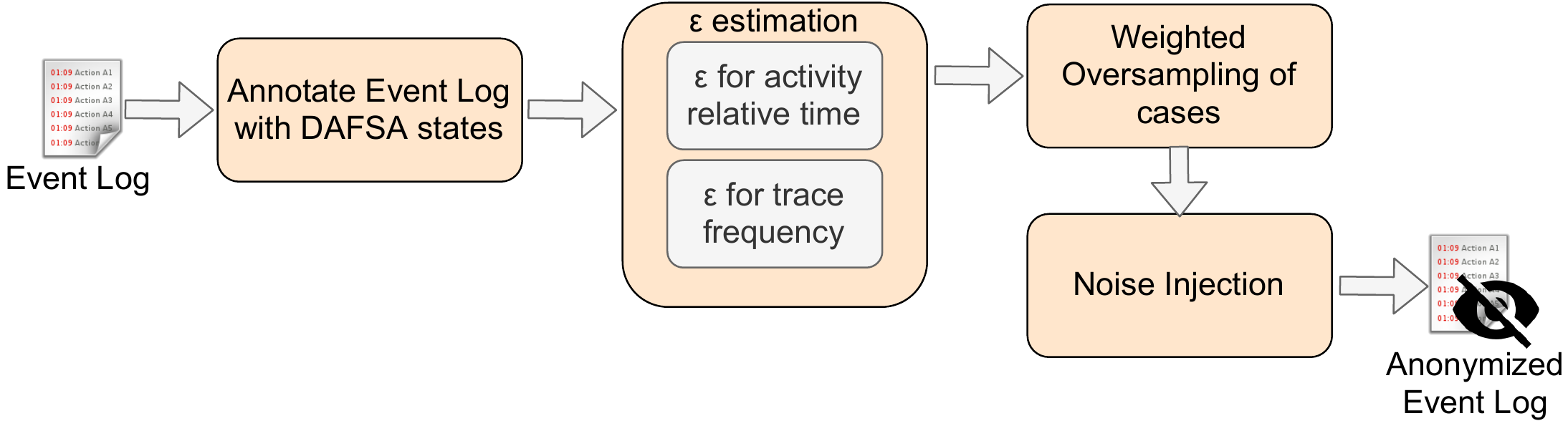}}
\vspace*{-6mm}
\caption{ Approach }
	\label{fig:approach}
  \vspace*{-7mm}
\end{figure}

Given the above definitions, this section translates the guessing advantage parameter $\delta$, into an algorithm that anonymizes the event log. Fig.~\ref{fig:approach} outlines 
the proposed approach. First, we construct the DAFSA state annotated event log as mentioned in Sect.~\ref{sec:event_log_rep}. We use the algorithm proposed by Reissner et al.~\cite{reissner2017scalable} to build the DAFSA. Second, we calculate the $\epsilon$ values needed by the approach. Finally, we use the calculated $\epsilon$ values to perform oversampling of cases and time noise injection into the event log. Below, we describe the steps in detail.

\subsection{ \texorpdfstring{$\mathcal\epsilon$}{Lg} Estimation}
\label{sec:app:eps}
Given the DAFSA annotated event log, we group the common prefixes and suffixes in traces by the DAFSA transitions. This grouping has two outputs: a distribution of execution time of activity instances that goes through the same DAFSA transitions and the DAFSA transitions contingency table. 
A case's time attribute comprises two components: the start time of the case and the relative execution time of every activity in the case (a time difference between its execution timestamp and the start timestamp of the case) in a time unit.

To calculate $\epsilon$ for the distribution of execution time instances, grouped by a DAFSA, we use Eq~(\ref{eqn:p_k}) and~(\ref{eqn:epsilon_time}), i.e., 
we use different $\epsilon$ values for every event in the event log. 
We normalize the input values to become in the range $[0,1]$ 
as mentioned in Sect.~\ref{sec:attack_model}. 
Eq~(\ref{eqn:epsilon_time}) provides the maximum $\epsilon$ to fulfill~\ref{int:req:time}.

On the other hand, the DAFSA transition contingency table contains a count histogram of the common prefixes and suffixes. To calculate the $\epsilon$ value for the marginals, we use Eq~(\ref{eqn:p_freq}) and~(\ref{eqn:epsilon_time}). The $\epsilon$ of all the DAFSA transitions of the contingency table is the same, but the noise is drawn for every transition independently. The propositions and the privacy proofs of $\epsilon$ estimation are presented in the supplementary material~\cite{supplementary}.

\subsection{ Weighted Oversampling of Cases}
\label{sec:app:oversample}

Given the above-calculated $\epsilon$ for trace frequency, we draw a random noise value from the Laplace distribution $Lap(\Delta f/\epsilon)$, for each DAFSA transition. Adding or removing a prefix/suffix of a trace to the log affects a single frequency count by $1$, so $\Delta f = 1$. We take the absolute value of the noise (additive noise) to fulfill~\ref{int:req:trace}. The noise (quantified by $\delta$) is added as increments to the frequencies of the contingency table. 
Using an absolute value (additive noise only), we alter the privacy guarantees of the differentially private mechanism, as the attacker can break Def.~\ref{def:udp} if the sampled noise falls within 
$[0,\Delta f]$. 
This happens with probability $1 - exp(-\epsilon/\Delta f)$. This additional risk is factored into $\delta$. 

Next, we increment the counts of the contingency table by means of oversampling of DAFSA transitions as defined in Def.~\ref{def:oversampling}. Finally, to maintain the consistency of the DAFSA transitions, we oversample the traces in the log using Def.~\ref{def:cases_oversampling}, as presented in Algorithm~\ref{alg:oversampling}.
This algorithm does not suppress nor add case variants to fulfill~\ref{int:req:trace}. The supplementary material~\cite{supplementary} provides the correctness proofs of Algorithm~\ref{alg:oversampling} (cf. Proposition 5, 6, and 7, and Theorem 1).

\SetNlSty{}{}{.}
\vspace*{-3mm}
\begin{algorithm}
\footnotesize 
\hspace*{\algorithmicindent} \textbf{Input:} Event Log, $\epsilon$, DAFSA  \\
 \hspace*{\algorithmicindent} \textbf{Output:}~Differentially~Private~Event~Log \\
 
 DafsaLookup $=$ Build DAFSA transitions to case variant lookup \;
 
 DafsaLookup[i].neededNoise$=|z_i|$, where $z_i$ is sampled from $Lap(\Delta f/\epsilon)$ independently for every transition $t_i$\;
 
 cnt = count(DafsaLookup .addedNoise $<$ DafsaLookup.neededNoise) \;
 
 \While{ cnt $>$ 0} {
  selectedTransition = pick a random transition such that DafsaLookup.addedNoise $<$ DafsaLookup.neededNoise\;
  
  pickedTraces = pick $x$ random traces that traverse selectedTransition, where $x$ = selectedTransition.neededNoise\;
  
  \ForEach {$t \in$  pickedTraces  } {
  
 DafsaLookup[t].addedNoise $++$ \;
  
  replicate a random case with a case variant $=$ t \;
 
  }
  cnt = count(DafsaLookup.addedNoise $<$ DafsaLookup.neededNoise)\;
  
 }
 generate new CaseID for every case\;
 shuffle cases \;
 
 \caption{Case Oversampling Algorithm}
 \label{alg:oversampling}

\end{algorithm}
\vspace*{-4mm} 


Algorithm~\ref{alg:oversampling} starts by constructing a correspondence (lookup) table between the DAFSA transitions and the case variants (line 3). This table maps every DAFSA transition to the case variants that traverse it. We use this lookup table to track the updates over transitions. 
Second, we draw a random noise from the Laplace distribution $Lap(\Delta f/\epsilon)$ (line 4) independently for every transition. Next, we count the DAFSA transitions that need noise injection and their needed noise (lines 5 and 6). Next, we pick a random transition that needs noise (with sampling weights of their occurrence frequency) (line 7). Then, we randomly choose a case variant that goes through the chosen transition (with sampling weights of their number of instances) (line 8). We replicate the chosen case variant by a number of times equals the needed noise (lines 9-11). For every replication, we choose a random case variant instance from the log to replicate. Next, we update the DAFSA lookup with the injected noise (line 10). We repeat this process until all the transitions have the minimum required noise. We keep $\epsilon$ values of the time attribute the same as the original cases to draw noise as in Sect.~\ref{sec:app:noise}. 
Then, we generate new case IDs for the cases (line 15), and we change their order (line 16).

\vspace*{-1mm}
\subsection{Noise Injection}
\label{sec:app:noise}
\vspace*{-1mm}
At this step, we have the DAFSA annotated event log, with the oversampled case instances and an $\epsilon$ value of each event. First, we divide the $\epsilon$ value of the replicated cases by the number of replications, as oversampling is considered repeating the same query more than once~\cite{dwork2014algorithmic}. Then, we draw a random noise from the Laplace distribution $Lap(\Delta f/\epsilon)$ to anonymize the relative time for every activity instance. Next, we transform the amount of noise from the normalized range ($[0,1]$) to the original range. Finally, we add the time noise to the original execution relative time.
After the noise injection, we transform 
the relative execution time of activities 
to timestamps. By the end of this step, the event log is anonymized by $\epsilon$ values calculated for the input maximum guessing advantage $\delta$.


\section{Evaluation}
\label{sec:eval}
\vspace*{-1mm}
To address the problem stated in Sect.~\ref{sec:intro} under requirements~\ref{int:req:trace} and~\ref{int:req:time}, the proposed method injects differentially private noise in two ways: (i) by oversampling some traces in the log; and (ii) by altering the event timestamps. The effect of the oversampling can be measured by means of an oversampling ratio, as defined below. Meanwhile, the alteration of timestamps leads to the anonymized log having a longer timespan than the original one, i.e., a time dilation. Below, we evaluate the method by studying the effect of its parameters ($\delta$ and $p$) on the time dilation and the oversampling ratio.

\textbf{Accuracy Measure.} We measure the time dilation by comparing the anonymized logs with the original event log as ground truth. We use the \textit{symmetric mean absolute percentage error} (SMAPE). SMAPE is the absolute difference between true value $t_i$ and the anonymized value $a_i$ divided by the sum of the absolute values of $t_i$ and $a_i$. $SMAPE=\frac {1}{n} \sum_{i=1}^{n} \frac{|t_i - a_i|}{|t_i| + |a_i|}$, where $n$ is the number of events in the event log.
We measure the effect on case variant frequencies by measuring the \textit{oversampling ratio}. To calculate the oversampling ratio, we divide the number of cases in the anonymized event log $m_a$ by the number of cases in the original event log $m$. $oversampling~ratio =\frac{m_a}{m}$.


\textbf{Event log Selection.} For our experiment, we use the real-life event logs publicly available at 4TU Center for Research Data\footnote{\url{https://data.4tu.nl/}} as of February 2021. 
The considered event log files, their characteristics, and descriptive statistics are available in the supplementary material~\cite{supplementary}.
The selected logs contain the process execution of different domains, e.g., government and healthcare.
From the set of available logs, we excluded the event logs that are not business processes (e.g., ``Apache Commons'', ``BPIC 2016'' logs, ``Junit 4.12"). 
Also, we exclude the set of event logs ``coSeLog WABO" as they are a pre-preprocessed version of BPI challenge 15. 
Finally, we select a single event log for each set of logs in BPI challenges 13, 14, 15, 17, 20.

\textbf{Running Environment.} We implement the proposed model as part of a prototype, namely Amun\footnote{\url{https://github.com/Elkoumy/amun}}. We use python 3.8.5 and PM4PY\footnote{\url{https://pm4py.fit.fraunhofer.de/}} for parsing the XES files. 
We run the experiment on a single machine with  AMD Opteron(TM) Processor 6276 and 32 GB memory. Also, 
we consider only the end timestamp to calculate the relative time of an event for simplicity, and the same approach is still valid to apply differential privacy.

In the following experiment, we keep the precision 
$p=0.1$.  We measure the activity execution time in hours. In a practical application, the user decides the time unit. We run the experiments ten times and report the averages. The supplementary material contains an experiment for different values of $p$.





	


\begin{table*}[hbtp]
\caption{ SMAPE dilation, Oversampling Ratio, and Runtime experiments}
\vspace*{-8mm}
\begin{center}
\scriptsize
	\begin{tabular}[t]{|c|c||c|c|c|c|c|c|c|c|c|c|c|c|c|c|}
\hline
\rotatebox[origin=c]{90}{Exp. Typ}
&$\delta$	& \rotatebox[origin=c]{90}{BPIC12}& \rotatebox[origin=c]{90}{BPIC13	}& \rotatebox[origin=c]{90}{BPIC14	}&\rotatebox[origin=c]{90}{BPIC15}
&\rotatebox[origin=c]{90}{BPIC17}&\rotatebox[origin=c]{90}{BPIC18}
&\rotatebox[origin=c]{90}{BPIC19}&\rotatebox[origin=c]{90}{BPIC20	}&\rotatebox[origin=c]{90}{CCC19}
&\rotatebox[origin=c]{90}{CreditReq}&\rotatebox[origin=c]{90}{Hospital}
&\rotatebox[origin=c]{90}{Sepsis	}&\rotatebox[origin=c]{90}{Traffic	}&\rotatebox[origin=c]{90}{	Unrine.}	\\\hline \hline
\multirow{5}{*}{\rotatebox[origin=c]{90}{SMAPE}}

&0.1&72.5&74.7&62.4&63.2&74.5&64.7&58.7&70.0&47.5&62.6&37.0&70.2&57.9&55.2\\
&0.2&67.7&70.4&59.5&60.6&69.9&61.8&52.7&66.4&46.3&62.6&36.6&66.4&57.8&51.8\\
&0.3&64.2&67.6&57.1&58.2&67.1&59.3&48.2&64.3&43.8&62.6&36.2&63.2&57.8&50.9\\
&0.4&61.3&65.5&55.1&56.2&64.9&57.3&44.5&62.7&44.2&62.6&35.8&60.6&57.8&50.1\\
&0.5&59.1&63.9&54.0&56.1&63.1&56.2&41.4&61.4&45.0&62.6&35.7&59.4&57.8&49.9\\

\hline \hline

\multirow{5}{*}{\rotatebox[origin=l]{90}{Oversample R}}&
0.1&2.5&1.9&3.8&13.9&1.6&7.1&1.2&2.0&14.7&1.0&14.4&5.2&1.0&1.1\\
&0.2&1.8&1.5&2.5&7.6&1.4&4.1&1.1&1.5&8.0&1.0&7.7&3.2&1.0&1.1\\
&0.3&1.6&1.3&2.1&5.4&1.3&3.2&1.1&1.4&5.7&1.0&5.5&2.6&1.0&1.0\\
&0.4&1.5&1.3&1.9&4.3&1.2&2.6&1.1&1.3&4.4&1.0&4.4&2.3&1.0&1.0\\
&0.5&1.4&1.2&1.7&3.7&1.2&2.3&1.1&1.3&3.9&1.0&3.6&2.1&1.0&1.0\\

 \hline \hline

\rotatebox[origin=c]{90}{Time}&
0.2&12.5&3.9&101.4&8.4&17.4&542.5&87.5&4.1&0.1&1.2&34.1&1.9&8.2&0.2\\\hline 

	\end{tabular}
\label{tbl:results}
\end{center}
\vspace*{-7mm}
\end{table*}

\textbf{Results.} The results are shown in Table~\ref{tbl:results}. We use 14 real-life event logs to study the method's parameters. 
The anonymized logs are available in the supplementary material~\cite{supplementary} with the estimated $\epsilon$'s for events. 
In this table, the oversampling ratio represents the amount of replicated cases to prevent singling out cases using the set of prefixes and suffixes of an activity. The larger the oversampling ratio, the more the cases will be replicated. For instance, the oversampling ratio for the event log CCC19 is large, because every case follows a different case variant. Similarly, BPIC15 has 1199 cases and 1170 case variants, which results in a large oversampling ratio. Meanwhile, all cases in the log Credit Requirement follow the same case variant, resulting in an oversampling ratio of just above 1. Likewise, the Traffic log has a small oversampling ratio as it contains 150~370 cases and only 231 case variants.

In our experiments, the reason for the significant dilation, measured in percentage SMAPE, is outliers in the event logs. For instance, the log BPIC3 contains only one case with a case duration of 2.1 years, and another case duration equals 1.3 years. This unique behavior causes the proposed approach to estimate smaller $\epsilon$ values to inject more noise to prevent case identification using their duration. Similarly, the log Sepsis cases contains outliers. 
Nevertheless, the histogram of case durations of the Hospital log does not show outliers, where the minimum number of cases with a unique case duration is 6. The same case happens with Unrineweginfectie and BPIC19 logs. Furthermore, as per the definition of Differential Privacy, we divide the $\epsilon$ by the number of replication. That results in a smaller $\epsilon$ value, which means more noise. 
For instance, the Sepsis cases event log has a large SMAPE percentage because the $\epsilon$ values are divided by the oversampling ratio of the replicated case, which results in more significant noise. 

\begin{figure*}
    \centering
    
       \subfigure[Sepsis Active cases over time]
    {
       \centering
         \includegraphics[width=1\columnwidth]{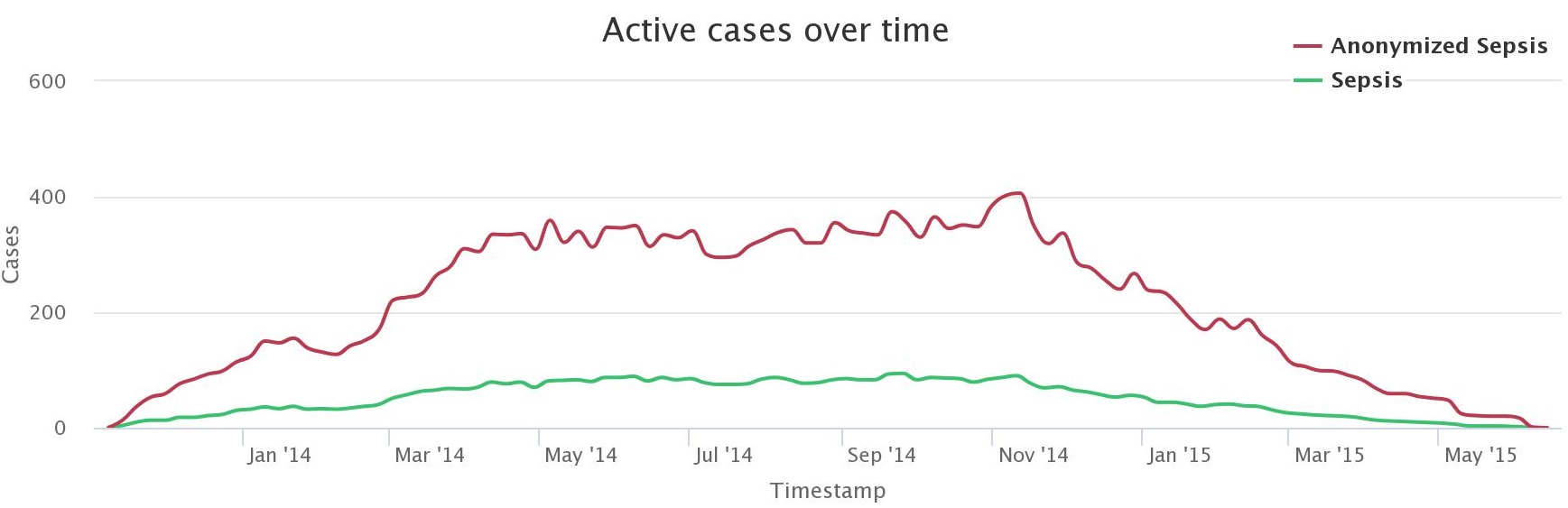}
          
         \label{fig:sepsis_cases_over_time}
        
    }~\subfigure[Traffic Active Cases over time]
    {
       \centering
         \includegraphics[width=1\columnwidth]{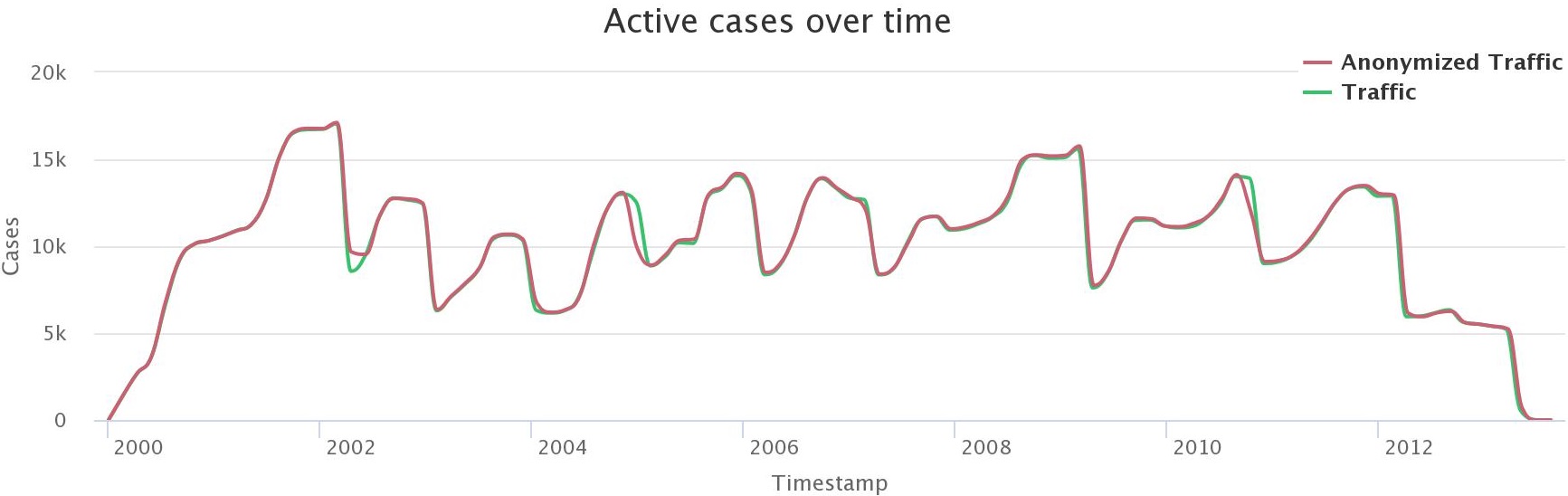}
         
         \label{fig:traffic_cases_over_time}
         
    }
    \\\vspace*{-4mm}
    \subfigure[Sepsis Case Variants Frequencies]
    {
       \centering
       
         \includegraphics[width=1\columnwidth]{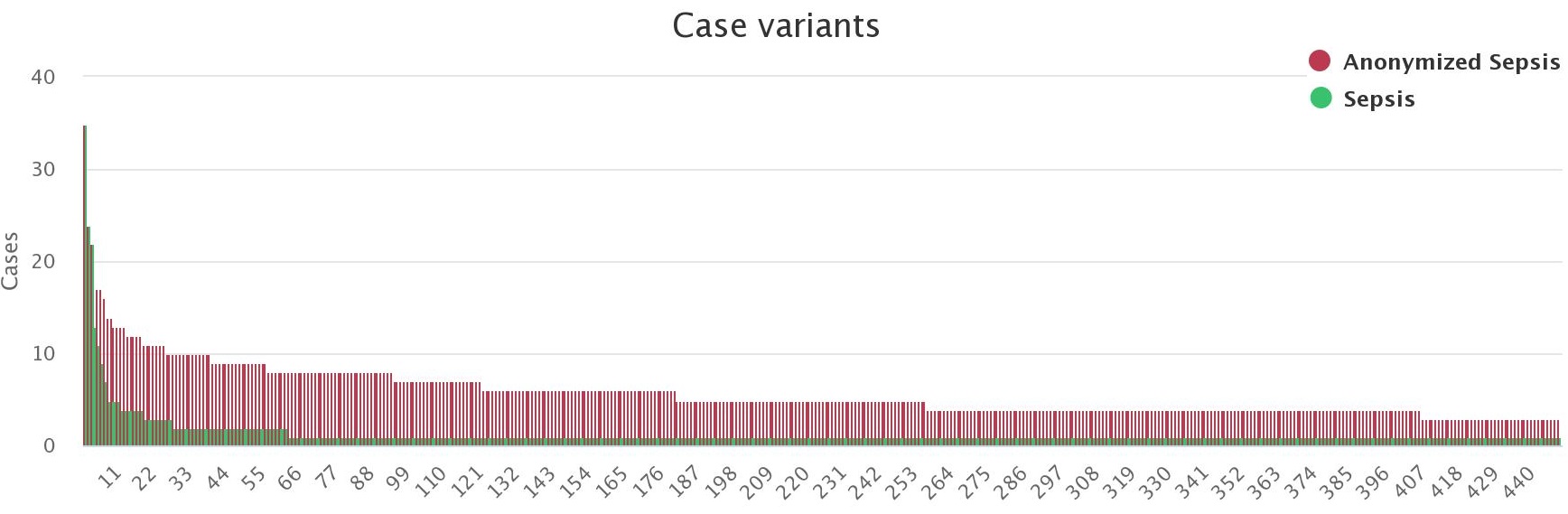}
         \label{fig:sepsis_case_variants}
         
    }~\subfigure[Traffic Case Variants Frequencies]
    {
       \centering
         \includegraphics[width=1\columnwidth]{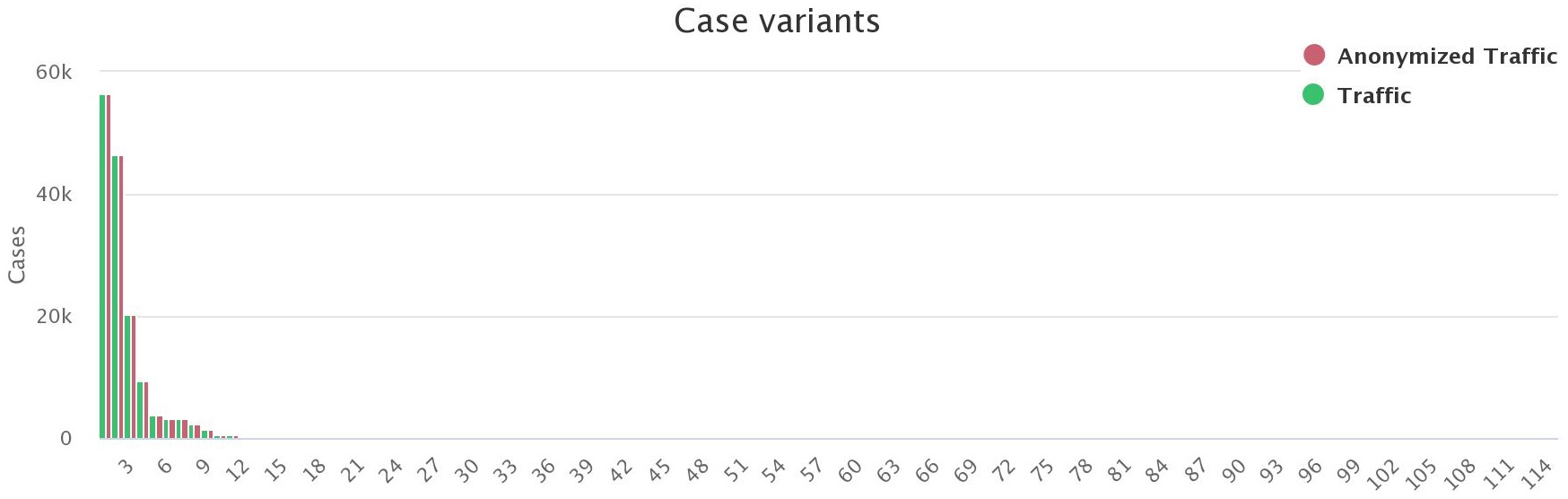}
         
         \label{fig:traffic_case_variants}
    }
    \vspace*{-4mm}
    \caption{Variant Analysis comparison between Traffic and Sepsis event logs and their anonymized versions, with $\delta=0.2$. The anonymized log is in red, and the original log is in green. The case variants are zoomed by 50\%. }
    \label{fig:variant_analysis}
    \vspace*{-5mm}
\end{figure*}

         
         
         
         
     

The variant analysis of both the Traffic and Sepsis cases event logs, and their anonymized version, with $\delta=0.2$ is shown in Fig.~\ref{fig:variant_analysis}. As reported in Table~\ref{tbl:results}, the oversampling ratio of the Traffic log is almost 1, and hence the active cases over time of both the Traffic log and its anonymized version are almost the same. However, there are time shifts due to the noise injected into the event execution time. Fig.~\ref{fig:traffic_cases_over_time} shows the active case over time, and Fig.~\ref{fig:traffic_case_variants} shows the frequency of case variants.  
Contrarily, the Sepsis cases log has a large sampling ratio; hence the anonymized version has more significant active cases over time, as shown in Fig.~\ref{fig:sepsis_cases_over_time} and Fig.~\ref{fig:sepsis_case_variants}. 

We conduct a wall-to-wall run time experiment to assess the efficiency of the method. We measure the time between reading the input XES file and the generation of its anonymized version. The results are reported in 
of Table~\ref{tbl:results}, and the values are in minutes. The run time grows with the growth of case variant (as it contains more DAFSA transition groups) and oversampling ratio (as it needs more oversampling iterations).
The reported execution times for logs with a large number of events and for low $\delta$ values are in the order of hours, e.g. 9 hours for BPIC18 (2.5 million events) with $\delta$=0.2. This is because the noise injection algorithm iterates multiple times over each transition (lines 6-14 in Algorithm~\ref{alg:oversampling}) and the number of DAFSA states for this log is high (638,242 states). This shortcoming can be tackled via parallelization, as the privacy quantification over each DAFSA transition is independent of others. The above experiments were all done using one single thread in order to avoid bringing additional variables (number of computing nodes and cores) into the experiments.

We acknowledge that the above observations are based on a limited population of logs (14). However, these logs were selected out of a broader population of close to 50 real-life logs based on justified selection criteria. 



\vspace*{-1mm}
\section{Conclusion and Future Work}
\label{sec:conclusion}
\vspace*{-2mm}
This paper proposed a concept of differentially private event log and a mechanism to compute such logs. A differentially private event log limits the increase in the probability that an attacker may learn a suffix of an individual's trace given a prefix (or vice-versa), or the timestamp of an activity in an individual's trace. To this end, we inject differentially private noise by oversampling the traces in the log. This approach neither suppresses nor adds case variants (cf. Sect.~\ref{sec:app:oversample}) and hence fulfills~\ref{int:req:trace}. To fulfill~\ref{int:req:time}, we quantify $\epsilon$ based on a technique that finds the maximum $\epsilon$ (minimum noise) that keeps the guessing advantage below $\delta$ (cf. Sect.\ref{sec:attack_model} and Prop.~\ref{prop:min_eps}). 

A limitation of the proposed method is that it anonymizes the relative time of each event with respect to the start of the case, but it does not anonymize the start times of the cases. We plan to address this limitation by applying methods for differentially private disclosure of time series. A second limitation is that the input log is assumed to have only three columns: case ID, activity label, and timestamp. Real-world event logs 
contain other columns, e.g., resources. To address this limitation, we need to extend the approach used to represent event logs, e.g., via multi-dimensional data structures as opposed to DAFSAs.

The proposed method introduces high levels of noise in the presence of unique traces or temporal outliers. To address this limitation, we plan to investigate an approach where high-risk traces are suppressed so that the amount of injected noise 
into the remaining traces is lower. Although suppression can significantly reduce the required noise level and strengthen the privacy guarantees, it breaks the property that the differentially private log has the same case variants as the original one. To mitigate this drawback, we will seek to define an approach to minimize the number of case variants that need to be suppressed, given the desired level of guessing advantage.

\vspace*{-1mm}
\medskip\noindent\textbf{Acknowledgments} Work funded by European Research Council (PIX project) and by EU H2020-SU-ICT-03-2018 Project No.
830929 CyberSec4Europe (\url{http://cybersec4europe.eu}). 
\vspace*{-2mm}
\bibliographystyle{IEEEtran} 

\bibliography{Amun2}

\begin{thebibliography}{10}
\providecommand{\url}[1]{#1}
\csname url@samestyle\endcsname
\providecommand{\newblock}{\relax}
\providecommand{\bibinfo}[2]{#2}
\providecommand{\BIBentrySTDinterwordspacing}{\spaceskip=0pt\relax}
\providecommand{\BIBentryALTinterwordstretchfactor}{4}
\providecommand{\BIBentryALTinterwordspacing}{\spaceskip=\fontdimen2\font plus
\BIBentryALTinterwordstretchfactor\fontdimen3\font minus
  \fontdimen4\font\relax}
\providecommand{\BIBforeignlanguage}[2]{{%
\expandafter\ifx\csname l@#1\endcsname\relax
\typeout{** WARNING: IEEEtran.bst: No hyphenation pattern has been}%
\typeout{** loaded for the language `#1'. Using the pattern for}%
\typeout{** the default language instead.}%
\else
\language=\csname l@#1\endcsname
\fi
#2}}
\providecommand{\BIBdecl}{\relax}
\BIBdecl

\bibitem{cohen2020towards}
A.~Cohen and K.~Nissim, ``Towards formalizing the gdpr’s notion of singling
  out,'' \emph{Proc. Natl. Acad. Sci.}, vol. 117, pp. 8344--8352, 2020.

\bibitem{rafiei2020tlkc}
M.~Rafiei, M.~Wagner, and W.~M. van~der Aalst, ``Tlkc-privacy model for process
  mining,'' in \emph{RCIS}.\hskip 1em plus 0.5em minus 0.4em\relax Springer,
  2020, pp. 398--416.

\bibitem{elkoumy2021privacy}
G.~Elkoumy, S.~A. Fahrenkrog-Petersen, M.~F. Sani, A.~Koschmider, F.~Mannhardt,
  S.~N. von Voigt, M.~Rafiei, and L.~von Waldthausen, ``Privacy and
  confidentiality in process mining--threats and research challenges,''
  \emph{arXiv preprint arXiv:2106.00388}, 2021.

\bibitem{dwork2014algorithmic}
C.~Dwork, A.~Roth \emph{et~al.}, ``The algorithmic foundations of differential
  privacy.'' \emph{Found. Trends Theor. Comput. Sci.}, vol.~9, pp. 211--407,
  2014.

\bibitem{lee2011much}
J.~Lee and C.~Clifton, ``How much is enough? choosing $\varepsilon$ for
  differential privacy,'' in \emph{Proc. ISC.}\hskip 1em plus 0.5em minus
  0.4em\relax Springer, 2011, pp. 325--340.

\bibitem{dwork2019differential}
C.~Dwork, N.~Kohli, and D.~Mulligan, ``Differential privacy in practice: Expose
  your epsilons!'' \emph{J. Priv. Confidentiality}, vol.~9, no.~2, 2019.

\bibitem{DumasRMR18}
M.~Dumas, M.~L. Rosa, J.~Mendling, and H.~A. Reijers, \emph{Fundamentals of
  Business Process Management, Second Edition}.\hskip 1em plus 0.5em minus
  0.4em\relax Springer, 2018.

\bibitem{daciuk2000incremental}
J.~Daciuk, S.~Mihov, B.~W. Watson, and R.~E. Watson, ``Incremental construction
  of minimal acyclic finite-state automata,'' \emph{Computational linguistics},
  vol.~26, no.~1, pp. 3--16, 2000.

\bibitem{laud2020framework}
P.~Laud, A.~Pankova, and M.~Pettai, ``A framework of metrics for differential
  privacy from local sensitivity,'' \emph{PoPETs}, pp. 175--208, 2020.

\bibitem{fahrenkrog2019pretsa}
S.~A. Fahrenkrog-Petersen, H.~van~der Aa, and M.~Weidlich, ``Pretsa: event log
  sanitization for privacy-aware process discovery,'' in \emph{ICPM}.\hskip 1em
  plus 0.5em minus 0.4em\relax IEEE, 2019, pp. 1--8.

\bibitem{mannhardt2019privacy}
F.~Mannhardt, A.~Koschmider, N.~Baracaldo, M.~Weidlich, and J.~Michael,
  ``Privacy-preserving process mining,'' \emph{BISE}, vol.~61, pp. 595--614,
  2019.

\bibitem{fahrenkrog2020pripel}
S.~A. Fahrenkrog-Petersen, H.~van~der Aa, and M.~Weidlich, ``Pripel:
  Privacy-preserving event log publishing including contextual information,''
  in \emph{BPM}.\hskip 1em plus 0.5em minus 0.4em\relax Springer, 2020, pp.
  111--128.

\bibitem{von2020quantifying}
S.~N. von Voigt, S.~A. Fahrenkrog-Petersen, D.~Janssen, A.~Koschmider,
  F.~Tschorsch, F.~Mannhardt, O.~Landsiedel, and M.~Weidlich, ``Quantifying the
  re-identification risk of event logs for process mining,'' in \emph{CAiSE},
  2020, pp. 252--267.

\bibitem{rafiei2020towards}
M.~Rafiei and W.~M. van~der Aalst, ``Towards quantifying privacy in process
  mining,'' \emph{arXiv preprint arXiv:2012.12031}, 2020.

\bibitem{batista2021uniformization}
E.~Batista and A.~Solanas, ``A uniformization-based approach to preserve
  individuals’ privacy during process mining analyses,'' \emph{Peer-to-Peer
  Networking and Applications}, pp. 1--20, 2021.

\bibitem{elkoumy2020secure}
G.~Elkoumy, S.~A. Fahrenkrog-Petersen, M.~Dumas, P.~Laud, A.~Pankova, and
  M.~Weidlich, ``Secure multi-party computation for inter-organizational
  process mining,'' in \emph{BPMDS}.\hskip 1em plus 0.5em minus 0.4em\relax
  Springer, 2020, pp. 166--181.

\bibitem{elkoumy2020shareprom}
G.~Elkoumy, S.~A. Fahrenkrog{-}Petersen, M.~Dumas, P.~Laud, A.~Pankova, and
  M.~Weidlich, ``Shareprom: {A} tool for privacy-preserving
  inter-organizational process mining,'' in \emph{BPM Demos}, 2020, pp. 72--76.

\bibitem{reissner2017scalable}
D.~Rei{\ss}ner, R.~Conforti, M.~Dumas, M.~La~Rosa, and A.~Armas-Cervantes,
  ``Scalable conformance checking of business processes,'' in \emph{OTM to
  Meaningful Int. Syst.}\hskip 1em plus 0.5em minus 0.4em\relax Springer, 2017,
  pp. 607--627.

\bibitem{barak2007privacy}
B.~Barak, K.~Chaudhuri, C.~Dwork, S.~Kale, F.~McSherry, and K.~Talwar,
  ``Privacy, accuracy, and consistency too: a holistic solution to contingency
  table release,'' in \emph{PODS}, 2007, pp. 273--282.

\bibitem{kifer2011no}
D.~Kifer and A.~Machanavajjhala, ``No free lunch in data privacy,'' in
  \emph{Proc. of ACM SIGMOD}, 2011, pp. 193--204.

\bibitem{supplementary}
\BIBentryALTinterwordspacing
G.~Elkoumy, A.~Pankova, and M.~Dumas, ``{Mine Me but Don't Single Me Out:
  Supplementary Material},'' Mar. 2021. [Online]. Available:
  \url{https://doi.org/10.5281/zenodo.5011364}
\BIBentrySTDinterwordspacing

\end{thebibliography}


\end{document}